\documentclass[%
reprint,
showpacs,
amsmath,amssymb,aps,
]{revtex4-1}

\usepackage{graphicx}
\usepackage{dcolumn}
\usepackage{bm}
\usepackage{color}

\newcommand{\md}{\mathrm d}
\newcommand{\mg}{\mu_{\rm g}}
\newcommand{\mgs}{\mu_{\rm g}^2}

\begin{document}


\title{The relation between the NJL model and QCD
with condensed gluons}%

\author{Hiroaki Kohyama}
\affiliation{Department of Physics,
National Taiwan University, Taipei 10617, Taiwan}

\date{\today}

\begin{abstract}
We try to find the relation between the three-flavor Nambu--Jona-Lasinio
model and QCD based on the hypothesis that the gluon momenta are sharply
condensed around the QCD scale, $\mg$. We find that the effective four-
and  six-fermion interactions, $G_4$ and $G_6$, should be scaled by
$G_4 \propto \mg^{-2}$ and $G_6  \propto \mg^{-5}$ being consistent
with the mass dimension counting in the obtained effective Lagrangian.
We then study the $\mg$ dependence on the phase diagram of the chiral
phase transition at finite temperature and chemical potential and the
location of the critical point. We find that the location of the critical point
are sensitively affected by the value of the introduced gluon energy scale.
\end{abstract}
\maketitle

\section{\label{sec:intro}%
Introduction}
Hadrons are composite objects constructed from quarks and gluons
via strong interaction. The first principle theory for quarks and
gluons is quantum chromodynamics (QCD), and one of our goals is
to describe observed properties of hadrons based on QCD.
Although the perturbative method is nicely adopted for high
energy phenomena, the investigation on the system at low energy,
such as composite bound state of hadrons, is still challenging 
issue. There we often employ chiral effective models with
four- and six-point interactions to study the hadron properties
at low energy.

The Nambu--Jona-Lasinio (NJL) model is one of successful
effective models of quantum chromodynamics (QCD) for describing
hadron physics~\cite{Nambu:1961tp}. The three-flavor version of
the model contains four- and six-fermion interactions in the
Lagrangian density, and the latter is called the
Kobayashi-Maskawa-'tHooft (KMT) term~\cite{Kobayashi:1970ji, 
'tHooft:1976up}. This KMT term explicitly breaks the ${\rm U_A}(1)$
symmetry which is not realized in the real world, and the model
incorporating the KMT term can describe the nonet meson properties
in a satisfactory manner (for reviews, see,
e.g.,~\cite{Vogl:1991qt, Klevansky:1992qe, Hatsuda:1994pi,
Rehberg:1995kh, Buballa:2005rept, Huang:2004ik}.)

Since the NJL model is regarded as an effective model for
QCD, the first principle theory of quarks and gluons, we believe
that the model should somehow be related to QCD. Recently,
the relation between the four-fermion interaction and the
original QCD Lagrangian has been discussed
in~\cite{Kohyama:2016dhd}, where it was found that the
two-flavor NJL Lagrangian can be connected to QCD with
hypothetical gluon condensate. It may also be interesting
to test whether the six-fermion interaction with the determinant
form suggested in \cite{Kobayashi:1970ji} can be derived based
on the same assumption that gluons are highly condensed
in the momentum space. Then, in this paper, we shall try to find
the relation between the effective six-fermion interaction
in the three flavor NJL model and  the original QCD
with the gluon condensate.

The paper is organized as follows. We start from the partition
function of QCD then apply the assumption of the gluon
condensate in Sec.~\ref{sec:qcd}. We then set the three-flavor
NJL model and perform the numerical analyses in Sec.~\ref{sec:njl}.
The discussions on the effective six-fermion interaction
and the concluding remark are given in Secs.~\ref{sec:remarks}
and \ref{sec:conclusion}.

\section{\label{sec:qcd}%
QCD with condensed gluons
}
Following the prescription employed in~\cite{Kohyama:2016dhd},
we first evaluate the partition function of QCD then introduce the
effect of gluon condensate for the sake of making the trial to find
the relation between the three-flavor NJL model and QCD.

\subsection{\label{subsec:QCD}%
QCD partition function}
We start from the following Lagrangian density
\begin{align}
 & \mathcal{L}_{\rm QCD}
   = \mathcal{L}_0 + \mathcal{L}_{\rm I}, \\
 & \mathcal{L}_0
   = \overline{q} (i\partial\!\!\!/ -m) q
      -\frac{1}{4}(\partial_\mu A^a_\nu - \partial_\nu A^a_\mu)^2,\\
 & \mathcal{L}_{\rm I}
      = g \overline{q} \gamma^\mu t^a q A^a_\mu 
      - gf^{abc}(\partial_\mu A_\nu^a) A^{\mu b}A^{\nu c} \nonumber \\
  & \qquad -\frac{1}{4} g^2
        (f^{eab}A^{a}_\mu A^{b}_\nu)(f^{ecd}A^{\mu c}A^{\nu d}),
\end{align}
where ${\mathcal L}_0$ and ${\mathcal L}_{\rm I}$ represent the
free and interacting parts. $q$ and $m$ are the quark field and
its current mass, $A_\mu^a$ is the gluon field, $g$ is the coupling
constant for the strong interaction and $t^a=\lambda^a/2$ with
$\lambda^a$ being the Gell-Mann matrices in the color space.

The partition function can be expanded by using the Taylor
series as
\begin{align}
  & \mathcal{Z}_{\rm QCD}
   =
    \int \! \mathcal{D}q \int \! \mathcal{D}\! A 
    \exp \left[
      i\int \md^4 x 
      \mathcal{L}_{\rm QCD}
       \right] \nonumber \\
   & =
      \int \! \mathcal{D}q \int \! \mathcal{D}\! A 
      e^{i\int \md^4 x{\mathcal L}_0} 
      \sum_{n=0}^{\infty}
      \frac{1}{n!}
      \left(  i \int \md ^4 x  \mathcal{L}_{\rm I} 
      \right)^n.
\end{align}
Here we study the terms up to the order of $g^4$ to consider
the four- and six-fermion interactions, then we see
\begin{align}
   &\mathcal{Z}_{\rm QCD}
     \simeq
      \int \! \mathcal{D}q \int \! \mathcal{D}\! A 
      e^{i\int \md^4 x{\mathcal L}_0} 
      \biggl[  1 
         +  \frac{1}{2} 
              \left( ig \int \md ^4 x  
                       {\mathcal L}_{\rm I}  \right)^2  \nonumber \\
   & \quad 
      + \frac{1}{3!}
           \left( ig \int \md ^4 x  
                  {\mathcal L}_{\rm I} \right)^3
      + \frac{1}{4!}
           \left( ig \int \md ^4 x  
                  {\mathcal L}_{\rm I} \right)^4 \,\,
      \biggr].
\label{eq:Z_QCD}
\end{align}
Note the first linear term for ${\mathcal L}_{\rm I}$ is not
required for our purpose, since it is expected to vanish.
Expanding and retaining the relevant terms, we have
\begin{align}
   &\mathcal{Z}_{\rm QCD}
     \simeq
      \int \! \mathcal{D}q \int \! \mathcal{D}\! A 
      e^{i\int \md^4 x{\mathcal L}_0} \nonumber \\
   & \quad \times
      \biggl[  1 + 
                 \frac{1}{2} \left( ig \int \md ^4 x  
                 \overline{q} \gamma^\mu t^a q A^a_\mu  
      \right)^2  \nonumber \\
   & \qquad \,\,
        + \frac{1}{3!} \cdot 3
            \left( ig \int \md ^4 x  
                    \overline{q} \gamma^\mu t^a q A^a_\mu  
            \right)^2  \nonumber \\
   & \qquad \quad \times
            \left( -i \frac{g^2}{4} \int \md ^4 y 
               (f^{hbc}A^{b}_\nu A^{c}_\rho)(f^{hde}A^{\nu d}A^{\rho e})
            \right)  \nonumber \\
   & \qquad \,\,
        + \frac{1}{4!} \cdot 4
            \left( ig \int \md ^4 x  
               \overline{q} \gamma^\mu t^a q A^a_\mu  
            \right)^3  \nonumber \\
   & \qquad  \quad \times
            \left( ig \int \md ^4 y  
               f^{bcd}(\partial_\nu A_\rho^b) A^{\nu c}A^{\rho d}  
            \right)
      \biggr].
\label{eq:Z_g4}
\end{align}
This is the partition function we will consider in what follows.

\subsection{\label{subsec:condensate}%
Treatment of the gluon condensate}
Having aligned the relevant terms for the partition function, we
now perform the functional integral of gluon under special
condition. As discussed in~\cite{Kohyama:2016dhd}, we assume
that the gluon momenta are condensed around the specific
scale, $p^2 \sim \mgs$. Then, for the gluon propagator of the
usual form,
\begin{align}
      \bigl\langle
           A^a_{\mu} (x)  
           A^b_{\nu} (y)
      \bigr\rangle
      =
      \int \frac{\md^4 p}{(2\pi)^4}
      \frac{-ig_{\mu \nu} \delta^{ab}}{p^2}
      e^{-i p \cdot (x-y)},
\end{align}
we apply the following replacement
\begin{align}
      \frac{1}{p^2}
      \to
      \frac{1}{\mgs}.
\label{eq:replace}
\end{align}
In more detail, $p^\mu \to \mg^\mu$ which becomes important
when we study the amplitudes at $g^4$ order. With the
replacement we obtain the Feynman rule shown below
\begin{align}
      \bigl\langle
           A^a_\mu (x)  
           A^b_\nu (y)
      \bigr\rangle
      =
      \frac{-ig_{\mu \nu} \delta^{ab} }{\mgs}      
      \delta^{(4)} (x-y)
\end{align}
after performing the momentum integration,
where the delta function induce the contact interaction
for fermion fields.

We should note that the propagator becomes infinite when
$y=x$, because the one-loop amplitude,
\begin{align}
  \phi_{\rm g}^{\rm bare}
  =  \int \frac{\md^4 p}{(2\pi)^4}
      \frac{-i}{p^2},
\label{eq:g_cond}
\end{align}
badly diverges. We then apply the renormalization
so as to obtain finite prediction, and we set
\begin{align}
      \bigl\langle
           A^a_{\mu} (x)  
           A^b_{\nu} (x)
      \bigr\rangle
      =
      g_{\mu \nu} \delta^{ab}
      \phi_{\rm g},
\end{align}
where $\phi_{\rm g}$ is finite renormalized quantity. Here
we call $\phi_{\rm g}$ as the gluon condensate being reminiscent
of the chiral condensate, $\phi_{\rm q} \equiv \langle \bar{q}q
\rangle$. One more attention should be paid when we consider the
six-fermion interaction, there we apply the following rule,
\begin{align}
 \int \frac{\md^4 p}{(2\pi)^4}
  \frac{ \bar{q} \mg \!\!\!\!\!\! / \,\,\, q }{\mgs} e^{-i(x-y)}
  \to 
  \frac{\bar{q} q}{\mg} \delta^{(4)}(x-y),
\end{align}
with the relation $\mg \!\!\!\!\!\! /\,\,\, q = \mg q$
set by using the Dirac equation $p\!\!\! / q = \sqrt{p^2} q$.

Based on the rules shown above, one can easily integrate
out the gluon degree of freedom,
\begin{align}
   &\mathcal{Z}_{\rm QCD}
     \simeq
      {\mathcal N}_A \int \! \mathcal{D}q 
      e^{i\int \md^4 x{\mathcal L}_0} \nonumber \\
   & \quad \times
      \biggl[  1 + 
        \frac{i g^2}{2 \mgs} 
        \int \md ^4 x  
        \left( \overline{q} \gamma^\mu t^a q  \right)  
        \left( \overline{q} \gamma_\mu t^a q  \right)  
        \nonumber \\
   & \qquad \,\,
        + \frac{i g^4}{\mu_{\rm g}^5} 
            \int \md ^4 x  \,\,
             f^{abc} ( \overline{q} \gamma_\mu t^a q )
   ( \overline{q} \gamma^\mu t^c q )
   ( \overline{q} t^b q ) 
      \biggr],
\label{eq:Z_g4_final}
\end{align}
where ${\mathcal N}_A$ is the over all constant relating to the
gluon functional integral, and we keep only the leading
contribution for each term. Moving back the resulting terms
inside the exponential using the relation
$1+\epsilon \simeq e^{\epsilon}$, we arrive at the following
effective Lagrangian density,
\begin{align}
   &\mathcal{L}_{\rm eff}
     =
     \overline{q} (i\partial\!\!\!/ -m) q
     + 
        \frac{g^2}{2 \mgs} 
        \left( \overline{q} \gamma^\mu t^a q  \right)  
        \left( \overline{q} \gamma_\mu t^a q  \right)  
        \nonumber \\
   & \qquad \,\,
        + \frac{g^4}{\mu_{\rm g}^5} 
             f^{abc} ( \overline{q} \gamma_\mu t^a q )
   ( \overline{q} \gamma^\mu t^c q )
   ( \overline{q} t^b q ).
\label{eq:Z_g4_2}
\end{align}
Thus we have obtained the four- and six-fermion contact
interaction based on the hypothesis of the condensed gluons.
It is interesting to note that the effective couplings 
are scaled by
\begin{equation}
  G_4 \propto \frac{g^2}{\mgs}, \quad
  G_6 \propto \frac{g^4}{\mg^5},
\end{equation}
as expected by the mass counting. It should
also be noted that if one sets $G_4$, $G_6$ to be constants
as usually done in practical model studies, the model
loses the renormalizability.

\subsection{\label{subsec:difficulty}%
Difficulty of deriving the KMT term}
We have discussed how the forms of the four- and six-fermion
contact interactions arise from the QCD with condensed gluons
in the previous subsection. Here it may be important to note that
the last term representing the six-fermion interaction vanishes
due to the antisymmetric property of $f^{abc}$. This indicates
that the ${\rm U_A}(1)$ breaking term does not appear through
the procedure based on the assumed gluon condensate.
We think this is natural consequence since the original QCD
Lagrangian is symmetric under the ${\rm U_A}(1)$ transformation.
Moreover, the fact that the KMT term includes the mixture of three
flavors, up, down and strange, can not be related with
our starting partition function since the original Lagrangian
does not include flavor structure. We will present further
discussions on this difficulty in Sec.~\ref{sec:remarks}.

\section{\label{sec:njl}%
The model}
We have checked the mass dimensions of the four- and six-point
couplings in the previous section through seeing the relation
between the effective couplings and the original QCD. We are now
going to study the effect of the condensed energy scale $\mg$ here.

\subsection{\label{subsec:njl}%
Three flavor NJL model}
For the sake of testing the $\mg$ dependence on the model
predictions, we consider the following effective Lagrangian
\begin{align}
   &\mathcal{L}_{\rm NJL}
     = \overline{q} (i\partial\!\!\!/ -m) q
     + 
        \frac{G_0}{\mgs} 
        \left[   ( \overline{q}  q  )^2 + (\overline{q}i\gamma_5  q  )^2 
        \right]
        \nonumber \\
   & \qquad \,\,
     + \frac{K_0}{\mu_{\rm g}^5} 
             [\det \bar{q}(1-\gamma_5)q + \det \bar{q}(1+\gamma_5)q],
\end{align}
with  the dimensionless couplings, $G_0$ and $K_0$. Where the effective
six-fermion term is introduced following \cite{Kobayashi:1970ji}, and the
determinant runs over the flavor space. If we write two couplings as
$G_0/\mgs =G$ and $K_0/\mg^5=K$, the model reduces to the usual
NJL model. After the mean-field approximation, we have the linearized
form,
\begin{align}
   &\mathcal{L}
     =
        \overline{q} (i\partial\!\!\!/ -M) q
     - \frac{2 G_0}{\mgs} 
        \left(  \phi_{\rm  u}^2 + \phi_{\rm d}^2 + \phi_{\rm s}^2  \right)
     + \frac{4 K_0}{\mu_{\rm g}^5} 
        \phi_{\rm  u} \phi_{\rm  d} \phi_{\rm s},
\end{align}
where $M$ represents the diagonal mass matrix for constituent 
quarks, $M_i = m_i -4G \phi_i + 2K \phi_j \phi_k 
(i \neq j \neq k)$, and $\phi_i$ indicate the chiral condensates,
$\phi_i \equiv \langle \bar{i}i \rangle$. The expectation
values of the chiral condensates are determined by the gap
equations,
\begin{align}
  \phi_i
  = -{\rm tr} \!\! \int \!\! \frac{\md^4q}{(2\pi)^4}\,
      \frac{i}{q \!\!\! / -M_i},
\end{align}
which is derived under the stational condition of the effective
potential ${\mathcal V} = -\ln Z /(V \beta)$ with the inverse
temperature $\beta = 1/T$. The above expression
quadratically diverges, then we will introduce the
three-momentum cutoff $\Lambda$ to obtain finite quantity.

The model has seven parameters, the three-momentum cutoff
$\Lambda$, the four- and six point couplings $G_0$, $K_0$,
the current quark masses $m_{\rm u}$, $m_{\rm d}$, $m_{\rm s}$,
and the gluon condensate scale $\mg$. 
Following~\cite{Hatsuda:1994pi}, we first set
$m_{\rm u}=m_{\rm d}=5.5$MeV,
then determine the four parameters
$\Lambda$, $G_0$, $K_0$, $m_{\rm s}$ by using the four physical
observables, $m_\pi=138$MeV, $f_\pi=92$MeV, $m_{\rm K}=495$MeV,
$m_{\eta^{\prime}}=958$MeV, at the scale $\mg=250$MeV.
The above condition leads the values, $\Lambda=631$MeV,
$G_0=0.288$, $K_0=0.09$, $m_{\rm s}=136$MeV, with
$m_{\rm u}=m_{\rm d}=5.5$MeV and $\mg=250$MeV.

\subsection{\label{subsec:pd}%
Phase diagram}
To see the effect of the gluon condensate on the chiral phase
transition, we draw the phase diagram on chemical potential
($\mu$)-temperature plane through changing the value of $\mg$
with the other parameters fixed. 
\begin{figure}[h!]
\begin{center}
   \includegraphics[width=7.5cm,keepaspectratio]{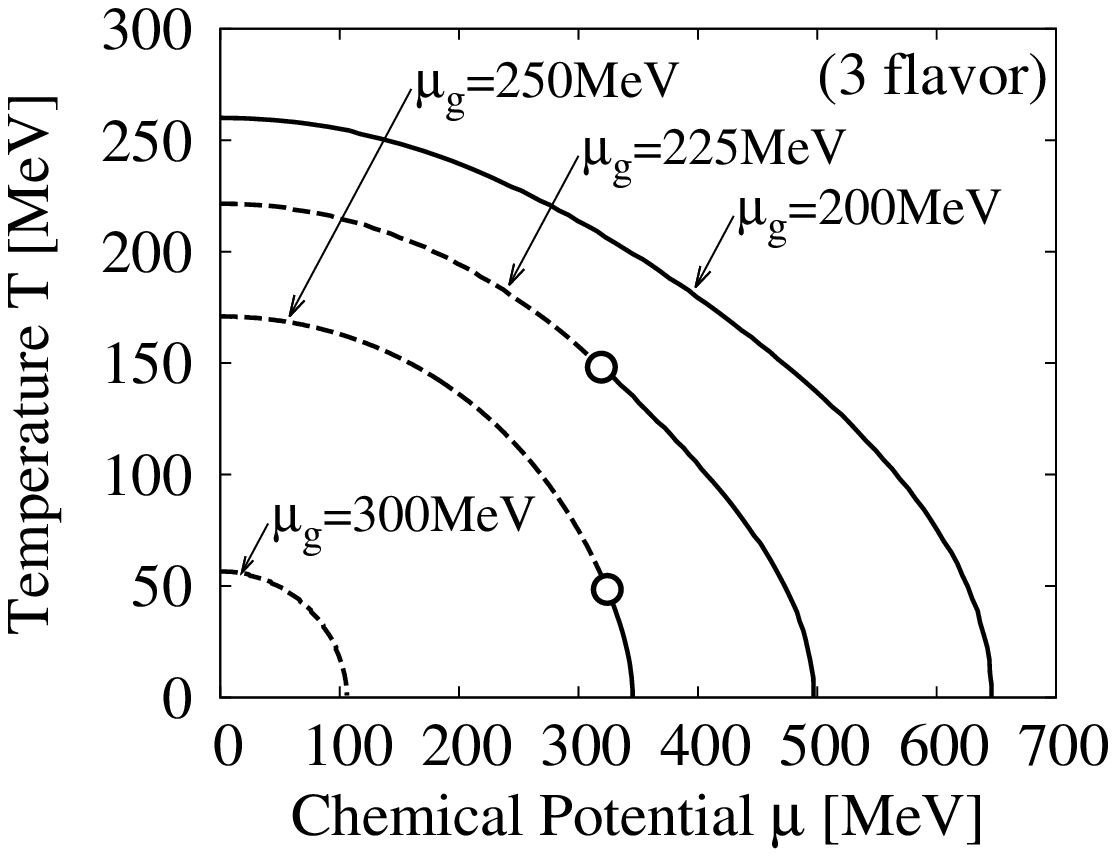}
   \includegraphics[width=7.5cm,keepaspectratio]{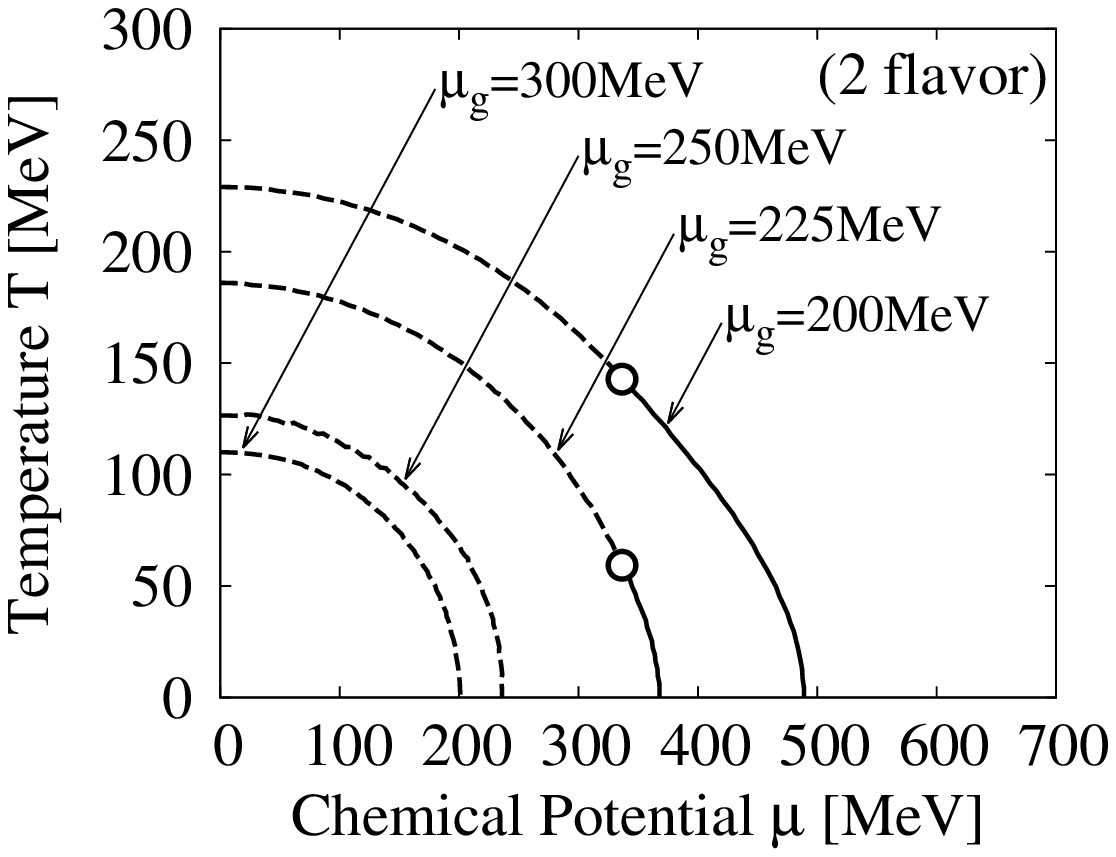}
   \caption{\label{fig:pd}
     $\mu_{\rm g}$ dependence on the phase diagram in the three- and
     two-flavor models. The solid (dashed) curves represent the first order
     (crossover) transition, and the circles do the critical points.
   }
\end{center}
\end{figure}
The upper panel of Fig.~\ref{fig:pd} displays the numerical results
on the phase diagram in the three flavor case for $\mg = 200$,
$225$, $250$ and $300$MeV. 
To study the effect of the six-fermion interaction, we also show
the phase diagram in the two-flavor model with basically the same
parameters $\Lambda=631$MeV, $G_0=0.288$ and
$m_{\rm u}=m_{\rm d}=5.5$MeV, but without the strange quark mass
and six-point interaction.
One sees that the region of the broken phase shrinks with
increasing with $\mg$; the tendency is common between
three- and two-flavor cases. This is easy to understand, since the
coupling strength for four-point interaction
$G_0/\mgs$ becomes smaller when one choose larger value of
$\mg$, then the symmetry tends to be restored at lower $T$ and
$\mu$. Quantitatively, the change of the area of the broken phase
on $\mu-T$ plane in the three-flavor model is more drastic than the
two-flavor case. This comes from the six-fermion interaction,
$K=K_0/\mg^5$ which is more sensitively affected by the energy
scale $\mg$; then the total change enhances in the three-flavor case.
We also note that the three-flavor case shows the stronger tendency
of the first order phase transition, which is also understood by the
above mentioned reasoning due to the six-fermion term. Thus the
investigation by the three-flavor case is important when one studies
the critical point, there the ${\rm U_A}(1)$ anomaly plays the crucial
role on the phase transition. We will perform more detailed analysis
on the location of the critical point in the following.

\subsection{\label{subsec:cp}%
Critical point}
\begin{figure}[h!]
\begin{center}
   \includegraphics[width=7.5cm,keepaspectratio]{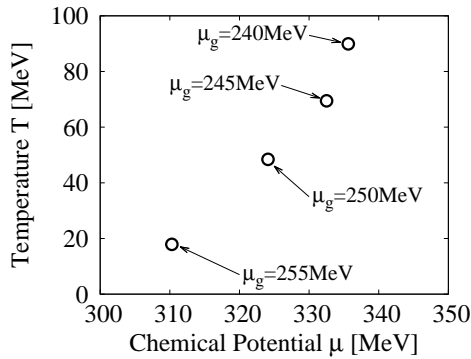}
   \caption{\label{fig:cp}
     $\mu_{\rm g}$ dependence on the critical point.
   }
\end{center}
\end{figure}

We see that the critical point drastically moves with changing the
gluon energy scale in the previous subsection.
It may also be interesting to discuss how the critical point moves
with varying the gluon condensate scale $\mg$ in more detail.
Figure~\ref{fig:cp} shows how the location of the critical point
$(T_{\rm cp}, \mu_{\rm cp})$ changes with respect to $\mg$.
One observes that the critical point moves towards the lower
temperature direction when the gluon energy scale becomes
larger. This is the straightforward consequence of the smaller $K$,
because the KMT term is intimately related to the strength of
the ${\rm U_A}(1)$ breaking and the term heightens the tendency
of the first order phase transition~\cite{Fukushima:2008wg}.
Thus the location or the existence of the critical pointe is indeed
sensitively affected by the value of the gluon condensate scale.

\section{\label{sec:remarks}%
Remark on the KMT term}
We have seen that the KMT term can not be obtained by staring
from the assumption of QCD with gluon condensate in Sec.~\ref{sec:qcd}.
We have also tried to obtain the term by following the discussion
in the paper~\cite{'tHooft:1976up} where the determinant term
with flavor structure is derived based on the ${\rm SU}(2)$ gauge theory
in the standard model (SM). However, we also faced the difficulty on the
derivation, since the extension from the ${\rm SU}(2)$
in the SM to the ${\rm SU}(3)$ in QCD contains a certain subtlety. In the SM case,
the Pauli matrices with respect to the gauge connection are set in the
flavor space, while in the QCD case the Gell-Mann matrices with respect 
to the gauge connection are defined in the color space. The difference is
crucial when one tries to obtain the determinant term in the flavor space;
in the QCD case the determinant structure can appear in the color space.
Thus, the extension from the ${\rm SU}(2)$ in the SM to the ${\rm SU}(3)$
in QCD is not straightforward with respect to the flavor indices,
so the derivation of the KMT term from the original QCD Lagrangian
is highly non-trivial.

\section{\label{sec:conclusion}%
Summary and conclusion}
We studied how the three-flavor NJL model can be related to the
QCD based on the hypothesis of condensed gluons in this paper.
There the correct mass dimensions on the effective four- and
six-fermion interactions, $G_4 \propto \mg^{-2}$ and
$G_6 \propto \mg^{-5}$, are found. We also found that it is not
possible to find the direct connection between the KMT term and
the QCD based on the assumption of the gluon condensate. We think
this is unavoidable
consequence since the original QCD Lagrangian does not include
any flavor structure in its form.

We then studied the gluon energy scale dependence on the phase
structure of the chiral phase transition. We see that the tendency
of the chiral symmetry breaking and the first order phase transition become
stronger when $\mg$ decreases. The tendency is easily understood
because the phenomenon of the symmetry breaking is expected to be
enhanced at low energy as the renormalization group analyses insists.

 Finally, we believe that the current analysis has a lot of applicability
 on physics relating to gluons since the gluon degree of freedom is
somehow incorporated to the effective model. Then we think it is
interesting to perform further investigations based on the QCD with
the gluon condensate.

\begin{acknowledgments}
The author thanks to T. Inagaki for discussions.
The author is supported by Ministry of Science and Technology
(Taiwan, ROC), through Grant No. MOST 103-2811-M-002-087.
\end{acknowledgments}

\end{document}